\documentclass[a4paper, amsfonts, amssymb, amsmath, reprint, showkeys, 
twoside, twocolumn,
elsarticle,
superscriptaddress]{revtex4-1}

\usepackage{amsthm}
\usepackage{mathtools}
\usepackage{physics}
\usepackage{xcolor}
\usepackage{graphicx}
\usepackage{hyperref}

\begin{document}
\title{Evidence of Coalescence Sum Rule in Elliptic Flow of Identified Particles in High-energy Heavy-ion Collisions}

\affiliation{Department of Physics and Astronomy, University of California, Los Angeles, California 90095, USA}
\affiliation{Key Laboratory of Nuclear Physics and Ion-beam Application (MOE), and Institute of Modern Physics, Fudan University, Shanghai-200433, People’s Republic of China}
\author{Amir Goudarzi}
\affiliation{Department of Physics and Astronomy, University of California, Los Angeles, California 90095, USA}
\author{Gang Wang}\email{gwang@physics.ucla.edu}
\affiliation{Department of Physics and Astronomy, University of California, Los Angeles, California 90095, USA}
\author{Huan Zhong Huang}
\affiliation{Department of Physics and Astronomy, University of California, Los Angeles, California 90095, USA}
\affiliation{Key Laboratory of Nuclear Physics and Ion-beam Application (MOE), and Institute of Modern Physics, Fudan University, Shanghai-200433, People’s Republic of China}

\begin{abstract}

The major goal of high-energy heavy-ion collisions is to study the properties of the deconfined quark gluon plasma (QGP), such as partonic collectivity.
The collective motion of constituent quarks can be derived from the anisotropic flow measurements of identified hadrons within the coalescence framework. Based on published results of elliptic flow ($v_2$),
we shall test the coalescence sum rule using $K^{\pm}$, $p$, ${\bar p}$, $\Lambda$ and ${\bar \Lambda}$, and further extract 
$v_2$ values for produced $u$($d$, $\bar u$, $\bar d$), $s$ and $\bar s$ quarks, as well as transported $u(d)$ quarks in 10-40\% Au+Au collisions at $\sqrt{s_{\rm NN}}=$ 7.7, 11.5, 14.5, 19.6, 27, 39 and 62.4 GeV. We also attempt to link the $v_2$ difference between $\pi^-$ and $\pi^+$ to the different numbers of $u$ and $d$ quarks in the initial gold ions, and to relate the $v_2$ measurements of multi-strange hadrons to the formation times of $\phi$, $\Omega^\pm$ and $\Xi^+$.

\end{abstract} 

\maketitle

\section{Introduction}
High-energy heavy-ion collisions have been performed at experimental facilities such as the Relativistic Heavy Ion Collider (RHIC) and the Large Hadron Collider (LHC), to create a deconfined partonic matter, quark gluon plasma (QGP), and to probe its properties.
One of the QGP signatures is partonic collectivity, in which scenario, the  spatial anisotropies of the initial  participating zone are converted by the pressure gradients of the QGP into the momentum  anisotropies  of  final-state  particles.
Accordingly,
experimental data analyses often express
the azimuthal distributions of
emitted particles with a Fourier expansion~\cite{Methods1,Methods2}
\begin{equation}
\frac{dN}{d\varphi}  \propto  1 + \sum_{n=1}^{\infty} 2v_n \cos[ n(\varphi -\Psi_{\rm RP})], 
\label{equ:Fourier_expansion}   \end{equation}
where $\varphi$ denotes the azimuthal angle of the particle
and $\Psi_{\rm RP}$ is the reaction plane azimuth (defined by the impact parameter  vector). The Fourier coefficients,  
\begin{equation}
v_n = \langle \cos [n(\varphi -\Psi_{\rm RP})] \rangle \,,  
\label{equ:Fourier_coefficient}
\end{equation}
are referred to as anisotropic flow of the $n^{\rm th}$ harmonic.
Here the average is taken over all particles and over all events. 
By convention,  $v_1$ and $v_2$ are called ``directed flow" and ``elliptic flow",  respectively.
They 
reflect the hydrodynamic response of the QGP fluid to the initial geometry  of the collision system~\cite{HYDRO_review}.

The coalescence  mechanism  empirically describes hadronization in heavy-ion collisions, 
assuming  quark and gluon constituents join into a hadron when they are close to each other in space and traveling with similar velocities~\cite{Coalescence1,Coalescence2}.
It is argued that  coalescence, rather than fragmentation of partons, dominates the emission of hadrons with transverse momentum ($p_T$) up to 5 GeV/$c$
in central high-energy heavy-ion collisions  ~\cite{PhysRevLett.90.202303}. Since particle spectra typically fall steeply with increasing $p_T$, features in the $p_T$-integrated yields are essentially dominated by the  coalescence dynamics. 
Besides its various manifestations in particle spectra~\cite{PhysRevLett.90.202303,AMPT_minijet, PhysRevC.79.044905, PhysRevC.92.054904, Charmed_Hadrons}, coalescence also plays a prominent role in anisotropic flow~\cite{Coalescence2,PhysRevC.69.051901}. Taking $\pi^+(u{\bar d})$ as an example, without loss of generality, we set  $\Psi_{\rm RP}$ to zero, and ignore the normalization:
\begin{widetext}
\begin{eqnarray}
v_n^{\pi^+} 
&=& \iiint d\varphi^ud\varphi^{\bar d}d\varphi^{\pi^+} \cdot \cos(2\varphi^{\pi^+})[1+2v_n^{u}\cos(2\varphi^u)][1+2v_n^{\bar d}\cos(2\varphi^{\bar d})]  
 \delta(\varphi^{\pi^+}-\varphi^u)\delta(\varphi^{\pi^+}-\varphi^{\bar d}) \nonumber \\
&=& \int d\varphi^{\pi^+} \cdot \cos(2\varphi^{\pi^+})[1+2v_n^{u}\cos(2\varphi^{\pi^+})][1+2v_n^{\bar d}\cos(2\varphi^{\pi^+})] \nonumber \\
&\approx& \int d\varphi^{\pi^+} \cdot \cos(2\varphi^{\pi^+})[1+(2v_n^{u}+2v_n^{\bar d})\cos(2\varphi^{\pi^+})] \nonumber \\
&=& v_n^{u}+v_n^{\bar d}.
\label{eq:coal}
\end{eqnarray}
\end{widetext}
The $\delta$ functions are there to enforce coalescence, and the approximation ignores the higher-order term, $v_n^u v_n^{\bar d}$,  assuming $v_n \ll 1$.
With the assumption of comoving quarks, Eq.~(\ref{eq:coal})
demonstrates the coalescence sum rule~\cite{v1}:  the $v_n$ of the resulting mesons or baryons is the summed $v_n$ of
their constituent quarks.
Note that in testing the coalescence sum rule with different hadron species, we only use the constituent quark properties at hadronization.
We cannot differentiate whether the quark collectivity originates from the hydrodynamic evolution or initial-state fluctuations.

If $n_q$ constituent quarks have the same $p_T$ and the same $v_n$, we have the number-of-constituent-quark (NCQ) scaling~\cite{NCQ0}:
\begin{equation}
    v_n^h(p_T^h) =n_q\cdot v_2^q(p_T^h/n_q) =n_q\cdot v_2^q(p_T^q).
\end{equation}
The approximate NCQ scaling of $v_2$ has been observed experimentally at RHIC and LHC energies~\cite{NCQ1,NCQ2,NCQ3,NCQ4,NCQ5,NCQ6,NCQ7}, evidencing partonic collectivity in these heavy-ion collisions.
While the same-$p_T$ condition is roughly another expression of ``traveling with similar velocities'' in the coalescence process, the same-$v_n$ requirement could be released to reveal anisotropic flow  of different quarks.
Significant $v_2$ differences have been discovered 
between particles and corresponding antiparticles at the RHIC Beam Energy Scan (BES)~\cite{NCQ6, Olivia}, and the observed $v_2$ orderings could be explained by a coalescence picture with  different $v_2$ values for (anti)quarks produced in pair and for $u$($d$) quarks transported from initial-state nuclei towards midrapidity ($y\approx0$)
~\cite{Dunlop}. 
Furthermore, $p_T$-integrated $dv_1/dy$ values for $K^{\pm}$, $p$(${\bar p}$) and $\Lambda$(${\bar \Lambda}$) have been used to test the coalescence sum rule for both produced quarks and transported quarks~\cite{v1}, and $v_1$ slopes have been extracted for produced $u$($d$, ${\bar u}$ and ${\bar d}$), $s$ and ${\bar s}$ quarks, as well as transported $u$($d$) quarks as functions of
collision energy~\cite{Gang}.
In the following, we shall extend this methodology to $p_T$-integrated $v_2$ with statistical uncertainties  much smaller than those of $v_1$. Therefore, the coalescence sum rule will be examined with better precision, and  within this framework, $v_2$ values will be estimated for constituent quarks in Au+Au collisions at BES.

\section{Coalescence sum rule}
\begin{figure}[t]
 \includegraphics[width=0.5\textwidth]{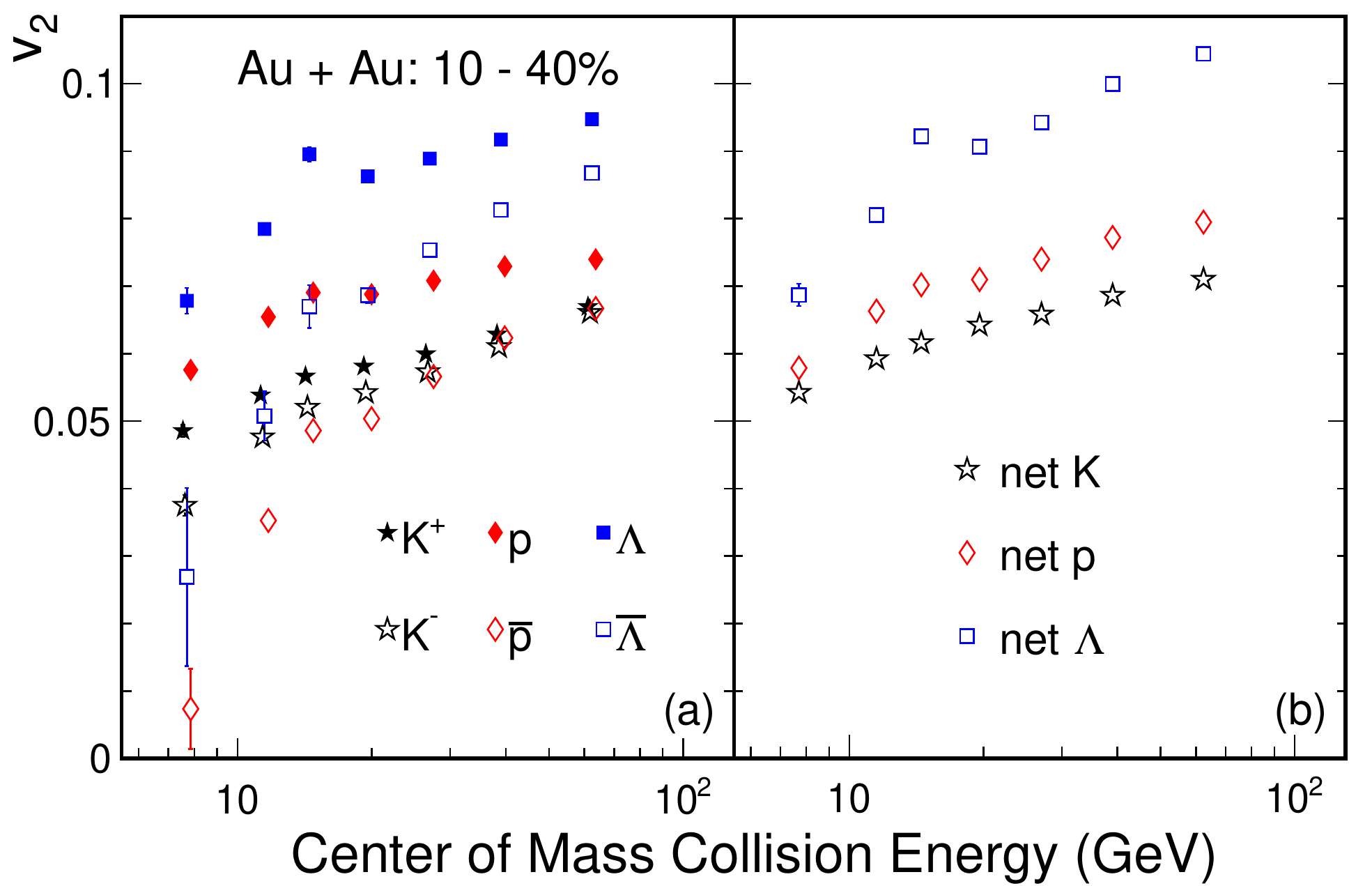}
 \caption{(Color online)
 Elliptic flow ($v_2$) for $K^{\pm}$, $p$, ${\bar p}$, $\Lambda$ and ${\bar \Lambda}$ (a),
 and for net $K$, net $p$ and net $\Lambda$ (b), in 10-40\% Au+Au collisions as function of beam energy, based on STAR data~\cite{Olivia}. Quoted errors are statistical uncertainties only.
 Some data points
are staggered horizontally to improve visibility.}
 \label{fig:m1}
\end{figure}

Figure~\ref{fig:m1}(a) illustrates the beam-energy dependence of $p_T$-integrated $v_2$ for $K^{\pm}$, $p$, ${\bar p}$, $\Lambda$ and ${\bar \Lambda}$ in 10-40\% Au+Au collisions, obtained from the corresponding $v_2(p_T)$ data published by the STAR Collaboration~\cite{Olivia}. The lower $p_T$ bound is 0.2 GeV/$c$ for $\pi^{\pm}$, $K^{\pm}$, $p$ and ${\bar p}$,
and 0.4 GeV/$c$ for all other hadrons, such that the majority of the particle yields are included, and the tracking efficiency correction can be largely ignored. The centrality range is selected in consideration of  {\it nonflow}.
The nonflow correlations~\cite{Nonflow} are  unrelated to the reaction plane orientation or the initial geometry, and  originate from transverse momentum conservation,  Coulomb and Bose-Einstein correlations,  resonance decays, inter- and intra-jet correlations, etc.
Although the pseudorapidity ($\eta$) gap of 0.05 between the particles of interest and the event plane suppresses some short-range nonflow contributions in the STAR measurements,
it is unlikely to eliminate
longer-range  correlations  due to, e.g., transverse momentum  conservation  and back-to-back  jet pairs. 
Nonflow is less influential in
the 10-40\% centrality, where $v_2$ itself is large,
and nonflow is diluted by multiplicity~\cite{STARv2}.

At each beam energy under study, $\Lambda$(${\bar \Lambda}$) shows larger $p_T$-integrated $v_2$ values than $p$(${\bar p}$), though they have very similar $v_2(p_T)$ functions~\cite{Olivia}.
This is because $\Lambda$(${\bar \Lambda}$) has higher mean $p_T$ values than $p$(${\bar p}$)~\cite{Ntrans,Strangeness}, and $v_2$ increases with $p_T$ for most of the accessible $p_T$ range. The $v_2$ difference between particles ($K^+$, $p$ and $\Lambda$) and their antiparticles ($K^-$, ${\bar p}$ and ${\bar \Lambda}$) warrants the effort to separate transported quarks and produced quarks.
The number of transported quarks is conserved, and transported quarks experience the whole system evolution.
In contrast, the total number of produced quarks is not conserved, and produced quarks are presumably created in different stages~\cite{Produced}. In experiments, produced quarks can be studied with purely ``produced'' particles, such as  $K^-$, $\bar p$ and $\bar \Lambda$, whereas transported quarks can be better probed with net particles that represent
the excess yield of a particle species over its antiparticle.
For example, 
\begin{equation}
v_2^{{\rm net}~ p} = (v_2^p - r_{{\bar p}/p} v_2^{\bar p})/(1-r_{{\bar p}/p}),
\end{equation}
where $r_{{\bar p}/p}$ is the ratio of observed $\bar{p}$ to $p$ yield at each beam energy. Similar expressions for net $K$ and net $\Lambda$ can be written by replacing $p$($\bar p$) with $K^+$($K^-$) and $\Lambda$($\bar \Lambda$), respectively. $v_2$ for net particles are shown in Fig.~\ref{fig:m1}(b), and will be  used later to examine the collectivity of transported quarks. 
 
\begin{figure}[b]
 \includegraphics[width=0.50\textwidth]{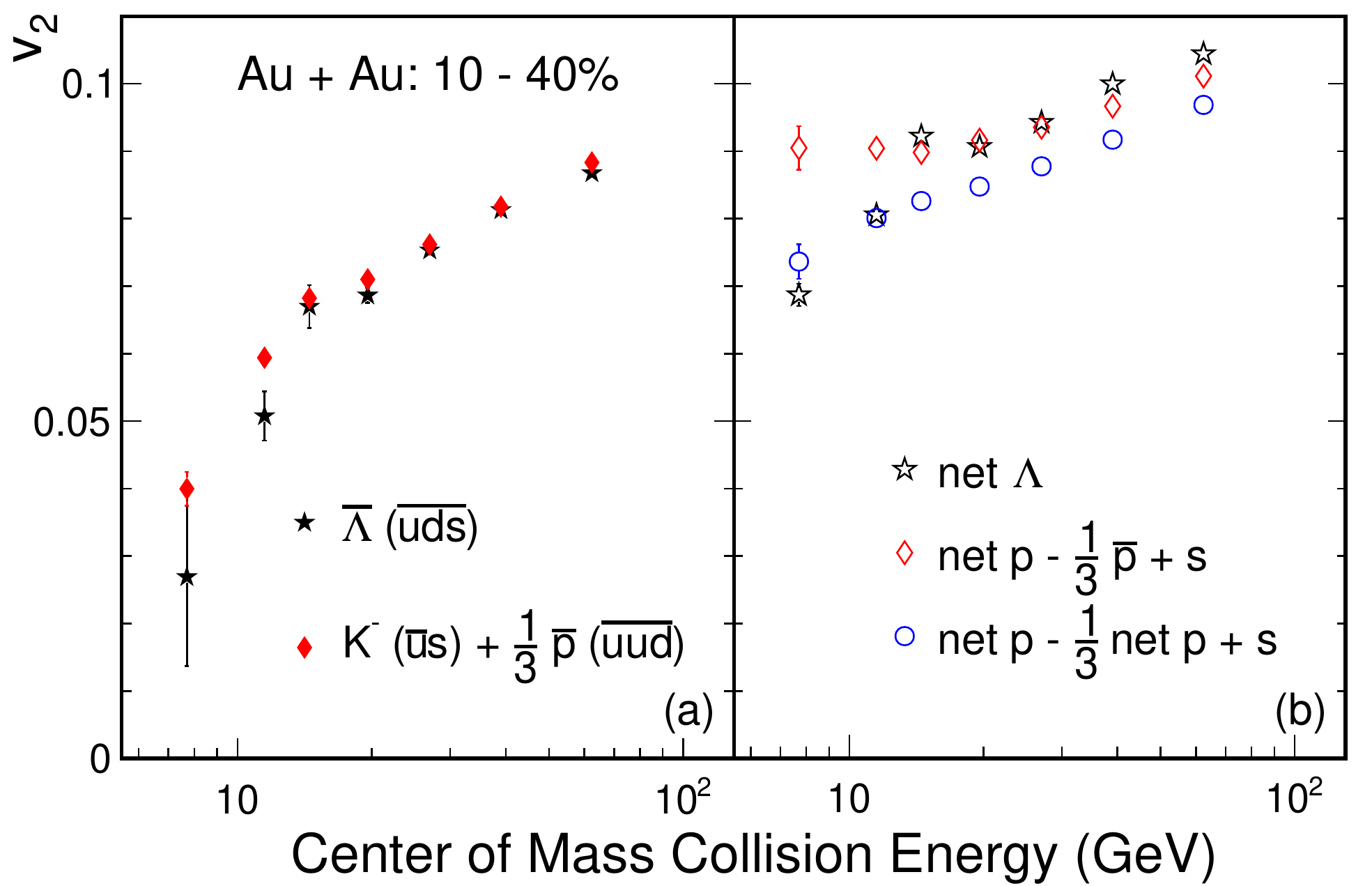}
 \caption{(Color online) $v_2$ versus beam energy for intermediate-centrality (10-40\%) Au+Au collisions. Panel (a) compares ${\bar \Lambda}$ with the prediction of the coalescence sum rule for produced quarks. Panel (b) shows two further sum-rule tests, based on comparisons with net $\Lambda$.}
 \label{fig:m2}
\end{figure}

The test of the coalescence sum rule for produced quarks is straightforward via produced particles. Figure 2(a) compares the observed $v_{2}$ for $\bar{\Lambda}(\bar{u}\bar{d}\bar{s})$ with the calculations for $K^{-}(\bar{u}s) + \frac{1}{3} \bar{p}(\bar{u}\bar{u}\bar{d})$. This comparison only involves  produced quarks, and we assume that $\bar{u}$ and $\bar{d}$ quarks have the same flow, and that $s$ and $\bar{s}$ have the same flow.
A close agreement appears at  $\sqrt{s_{\rm NN}} = 14.5$ GeV and above, supporting the coalescence picture. The possible deviations at 7.7 and 11.5 GeV are not statistically significant, but if confirmed, they would   imply the failure of one or more of the aforementioned
assumptions.
For example, if
the hadronic interactions 
dominate over partonic ones,
then coalescence is invalidated. Also, the degeneracy between $s$ and  $\bar s$ quarks may by lifted by the associated strangeness production,
$pp \rightarrow  p\Lambda(1115)K^+$~\cite{Associated}, whose role has also been manifested in other physics observables~\cite{v1,OffDia}.

The test of the collective behavior of $u$ and $d$ quarks is complicated by the extra component of transported quarks. In the limit of high $\sqrt{s_{\rm NN}}$, most $u$ and $d$ are produced, whereas in the limit of low $\sqrt{s_{\rm NN}}$, most of them are presumably  transported.
Transported quarks are more concentrated in net particles  than in particles, roughly in proportion to $N_{\rm particle}/N_{\rm net~particle}$~\cite{v1}. Therefore we employ net-$\Lambda$ and net-$p$ $v_2$ in these
tests. 
Figure~\ref{fig:m2}(b) compares net-$\Lambda$($uds$) $v_2$ with calculations in two scenarios.
The first (red diamond markers)  consists of net $p$($uud$) minus $\bar{u}$ plus $s$, with $\bar{u}$ estimated from $\frac{1}{3} \bar{p}$, and $s$ from $K^{-}(\bar{u}s) - \bar{u}$. Here we assume that a produced $u$ quark in net $p$ is replaced with an $s$ quark.
This sum-rule calculation agrees reasonably well with the net-$\Lambda$ data at $\sqrt{s_{\rm NN}} = 14.5$ GeV and above, and deviates significantly at 7.7 and 11.5 GeV.

The second calculation in Fig.~\ref{fig:m2}(b) corresponds to $\frac{2}{3}$
net $p$ plus $s$ (blue circle markers). In this scenario, we assume that the constituent quarks
of net $p$ are dominated by transported quarks in
the limit of low beam energy, and that one of the transported quarks is replaced with $s$. This approximation seems to work well at 7.7 and 11.5 GeV, and breaks down as the beam energy increases, with  disagreement between the black stars and blue circles above 11.5 GeV.
Overall, at each beam energy, net-$\Lambda$ $v_2$ can always be explained by one of the two coalescence scenarios. This suggests that the coalescence sum rule is valid for all the beam energies under study, as long as the difference between different quark species is taken into account.

In the published elliptic-flow data for heavy-ion collisions at top RHIC energies and at LHC energies, 
particles and antiparticles are rarely separated, which prevents us from performing the same test as in Fig.~\ref{fig:m2}.
The ratio $r_{{\bar p}/p}$ is about $0.8$ in Au+Au at 200 GeV~\cite{STAR_pbarp}, reflecting a sizeable transported-quark contribution to the particle production. At LHC energies, $r_{{\bar p}/p}$ is consistent with unity~\cite{ALICE_pbarp}, and we assume that all the quarks are produced in pair in the high-$\sqrt{s_{\rm NN}}$ limit. Therefore, the coalescence sum rule can still be tested with the $v_2$ values averaged over particles and antiparticles.
For example, based on ALICE data in 10-40\% Pb+Pb collisions at 2.76 TeV~\cite{NCQ7}, the $p_T$-integrated $v_2$ is $(8.06\pm0.06)\%$ for $p(\bar p)$, and $(9.71\pm0.04)\%$ for $K^\pm$, leading to  $(12.40\pm0.05)\%$ for $K^\pm + \frac{1}{3}p(\bar p)$, while the measured $v_2$ for $\Lambda(\bar \Lambda)$ is $(12.45\pm0.05)\%$, consistent with the coalescence expectation.

\section{Elliptic flow for quarks}
\begin{figure}[b]
 \includegraphics[width=0.50\textwidth]{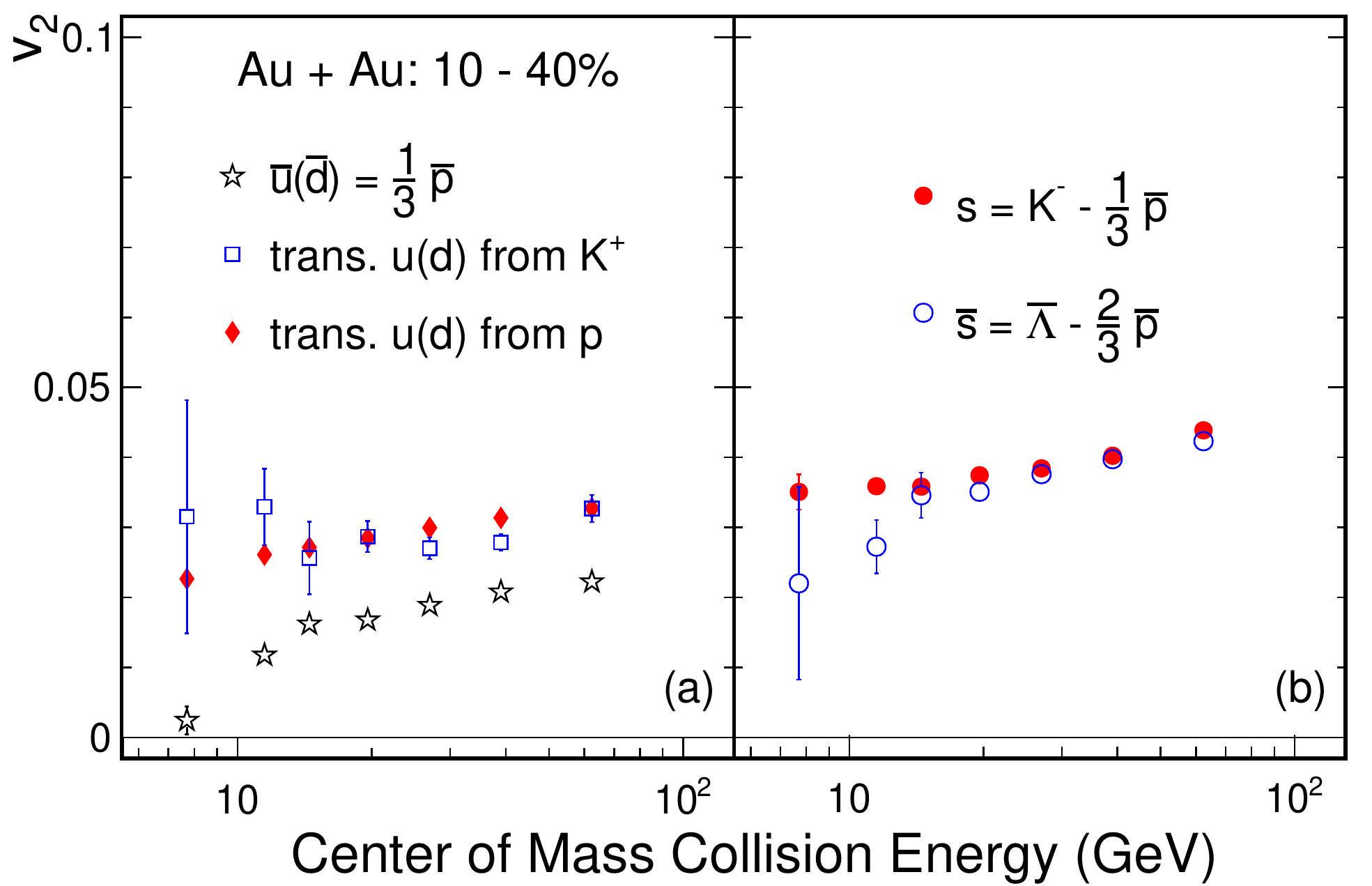}
 \caption{(Color online) $v_2$ for produced ${\bar u}$(${\bar d}$) and transported $u$($d$) quarks (a), and for $s$ and ${\bar s}$ quarks (b), in 10-40\% Au+Au collisions as function of
 beam energy.}
 \label{fig:m4}
\end{figure}

Assuming the coalescence sum rule, we shall further extract elliptic flow  for constituent quarks. 
$v_{2}$ for produced $u(d, \bar{u}, \bar{d})$ quarks is approximated with $\frac{1}{3} \bar{p} ({\bar u}{\bar u}{\bar d})$, as displayed in Fig.~\ref{fig:m4}(a). With diminishing beam energy, $p_T$-integrated $v_2$ of produced $u(d, \bar{u}, \bar{d})$ quarks becomes smaller, which could be partially explained by the gradually decreasing mean $p_T$. However, from 11.5 to 7.7 GeV, although mean $p_T$ only drops by less than relative $5\%$, $v_2$ quickly approaches zero,
indicating a dramatic change in physics.
As shown in Fig.~4 of Ref~\cite{Olivia}, in 10-40\% Au+Au collisions at $\sqrt{s_{\rm NN}}=7.7$ GeV, the $v_2$ value for $\bar p$ is slightly negative and consistent with zero for $p_T < 1$ GeV/$c$, and then increases with $p_T$ to a sizeable amount.
It is debatable whether this feature is due to a lack of the QGP state. At 7.7 GeV, the ratio $r_{{\bar p}/p}$ is only around $0.5\%$, so even if a low-$p_T$ $\bar p$ materializes from a QGP and carries a finite $v_2$, it is likely to be absorbed by the flowing baryon-rich environment. Therefore, the smallness of $v_2$ does not necessarily signify the disappearance of the QGP. In the following discussion, we still assume that the coalescence sum rule is valid for all the collision energies under study.

The collective behaviors of $s$ and $\bar{s}$  quarks are estimated from $K^{-}(\bar{u}s) - \bar{u}$ and  $\bar{\Lambda} (\bar{u}\bar{d}\bar{s}) - 2  \bar{u}$, respectively, as depicted in Fig.~\ref{fig:m4}(b).
In general, $s$ and $\bar{s}$  quarks show larger $p_T$-integrated $v_2$ values than produced $u(d, \bar{u}, \bar{d})$ quarks, for the same reason why  $\Lambda$(${\bar \Lambda}$) acquires larger $p_T$-integrated $v_2$ values than $p$(${\bar p}$): the former bears higher mean $p_T$ values.
The consistency between $s$ and $\bar s$ holds well at 14.5 GeV and above, and a seeming (but insignificant) split appears at 7.7 and 11.5 GeV.
The aforementioned associated strangeness production effectively converts the excess of $p$ (over $\bar p$) into $\Lambda(uds)$ and $K^+(u{\bar s})$. The mass difference between $\Lambda$ and $K^+$ will be reflected in the amount of collectivity they inherit from $p$, and causes asymmetry between $s$ and $\bar s$ quarks inside them. 
Later, $\Lambda$ and $K^+$ are melted in the QGP, and then $s$ and $\bar s$ quarks participate in the coalescence to form other strange hadrons. This intuitive picture not only expects
the $v_2$ difference between $s$ and $\bar s$, but also predicts a larger effect at lower energies, where baryon chemical potential is higher. 

Elliptic flow for transported quarks can be obtained with several approaches: by removing $\bar s$ and produced $u$ from $K^+(u{\bar s})$, and 
by removing produced $u(d)$ quarks from $p$ or net $p$. 
\begin{eqnarray}
v_{2}^{{\rm trans.} u(d)} &=&
[v_{2}^{K^+} - v_2^{\bar s} - (1 - f_{u}) \cdot v_{2}^{ \bar{u}}]/f_{u}
\\
&=&
[v_{2}^p/3 - (1 - f_{u(d)}) \cdot v_{2}^{ \bar{u}(\bar{d})}]/f_{u(d)}
\\
&=& \frac{ v_{2}^{{\rm net}~p} - (3 - N_{{\rm trans.} u + d}^{{\rm net}~p}) \cdot v_{2}^{ \bar{u}(\bar{d})}}{N_{{\rm trans.} u + d}^{{\rm net}~p}} ,
\end{eqnarray}
where $f_{u(d)}$ represents the fraction of transported $u$($d$) in all $u$($d$) quarks, and $N_{{\rm trans.} u + d}^{{\rm net}~p}$ is the number of transported quarks per net $p$, 
\begin{eqnarray}
N_{{\rm trans.} u + d}^{{\rm net}~p} &=& (2f_u N_p + f_d N_p)/(N_p - N_{\bar p}) \nonumber \\
&=& (2f_u + f_d)/(1-r_{{\bar p}/p}).
\label{eq:Nt}
\end{eqnarray}
Following Boltzmann statistics, we have
\begin{equation}
f_{u(d)} =\frac{ e^{{\mu_{u(d)}}/T_{\rm ch}}-e^{-{\mu_{u(d)}}/T_{\rm ch}}}{e^{{\mu_{u(d)}}/T_{\rm ch}}} = 1-e^{-{2\mu_{u(d)}}/T_{\rm ch}},
\label{eq:frac}
\end{equation}
where $\mu_{u(d)}$ is chemical potential for $u(d)$ quarks, and $T_{\rm ch}$ is chemical freeze-out temperature.
In the current scope, we always ignore the difference between $u$ and $d$, and take $\mu_{u(d)} = \mu_{\rm B}/3$, where $\mu_{\rm B}$ is baryon chemical potential.
Therefore, Eq.(\ref{eq:Nt}) becomes~\cite{Jinfeng}
\begin{equation}
N_{{\rm trans.} u + d}^{{\rm net}~p} = 3[1-e^{-2\mu_{\rm B}/(3T_{\rm ch})}]/(1-r_{{\bar p}/p}).    
\end{equation}
A similar idea as Eq.~(\ref{eq:frac}) maintains that $r_{{\bar p}/p}$ is roughly $e^{-2\mu_{\rm B}/T_{\rm ch}}$. Hence, in the limit of low $\sqrt{s_{\rm NN}}$ or high $\mu_{\rm B}$, $N_{{\rm trans.} u+d}^{{\rm net}~p}$ is close to three, whereas in the limit of high $\sqrt{s_{\rm NN}}$ or low $\mu_{\rm B}$, $N_{{\rm trans.} u+d}^{{\rm net}~p}$ approaches unity. These features are confirmed by Tab.~\ref{tab:Ntrans} that lists $f_{u(d)}$ and $N_{{\rm trans.} u+d}^{{\rm net}~p}$ as functions of $\sqrt{s_{\rm NN}}$ in 10 - 40\% Au+Au collisions, based on STAR data of  $\mu_{\rm B}$ and $T_{\rm ch}$~\cite{Ntrans}. 
\begin{table}
\caption{Chemical freeze-out parameters ($\mu_{\rm B}$ and $T_{\rm ch}$) for Strangeness Canonical Ensemble~\cite{Ntrans}, $f_{u(d)}$ and $N_{{\rm trans.} u+d}^{{\rm net}~p}$ in 10-40\% Au+Au collisions at RHIC BES. Errors in
parenthesis are systematic uncertainties. The values at 14.5 GeV are interpolated using results from other beam energies.}
\begin{tabular}{ c || c | c| c | c}
\hline
$\sqrt{s_{\rm NN}}$ (GeV) & $\mu_B$ (MeV) & $T_{\rm ch}$ (MeV) & $f_{u(d)}$ &$N_{{\rm trans.} u+d}^{{\rm net}~p}$ \\
\hline
7.7 & 387.2 (4.8) & 146.1 (1.2) & 0.829 & 2.50 
\\
\hline
11.5 & 282.2 (4.9) & 155.9 (1.6) & 0.701 & 2.16 
\\
\hline
14.5 & 234.1 (5.0) & 160.0 (1.9) & 0.623 & 1.97 
\\
\hline
19.6 & 182.0 (4.9) & 162.9 (2.3) & 0.525 & 1.76 
\\
\hline
27 & 135.3 (4.3) & 161.7 (2.5) & 0.428 & 1.57 
\\
\hline
39 & 98.8 (4.0) & 162.1 (2.8) & 0.334 & 1.41
\\
\hline
62.4 & 63.3 (3.7) & 160.7 (2.8) & 0.231 & 1.22 
\\
\hline
\end{tabular}
\label{tab:Ntrans}
\end{table}

The $v_2$ values for transported quarks from  $K^+$ and $p$ (or net $p$) corroborate each other, as exhibited in Fig.~\ref{fig:m4}(a). Results from $p$ and net $p$ are in good agreement, with different statistical uncertainties owing to our conservative error propagation. We opt to present the points with the smaller error bars in Fig.~\ref{fig:m4}(a). 
Ref~\cite{Dunlop} argues that the $v_2$ orderings between particles and antiparticles observed at RHIC BES
stem from the $v_2$ difference between transported and produced quarks.
Supposedly, transported quarks
undergo the entire evolution, receive more scatterings, and thus attain larger $v_2$ values. This speculation is supported by our $v_2$ extraction for constituent quarks.
Unlike $dv_1/dy$ for transported quarks that reveals a clear non-monotonic dependence on beam energy, evidencing the first-order phase transition~\cite{Gang}, $v_2$ for transported quarks does not demonstrate an apparent non-monotonic trend.
A possible explanation is that $v_1$ for transported quarks is mostly imparted at the early stage after the initial impact,
whereas $v_2$ for transported quarks is built up gradually by the pressure gradient of the QGP. Therefore, $v_1$ and $v_2$ have different sensitivities to the pertinent physics. 

\begin{figure}[t]
 \includegraphics[width=0.50\textwidth]{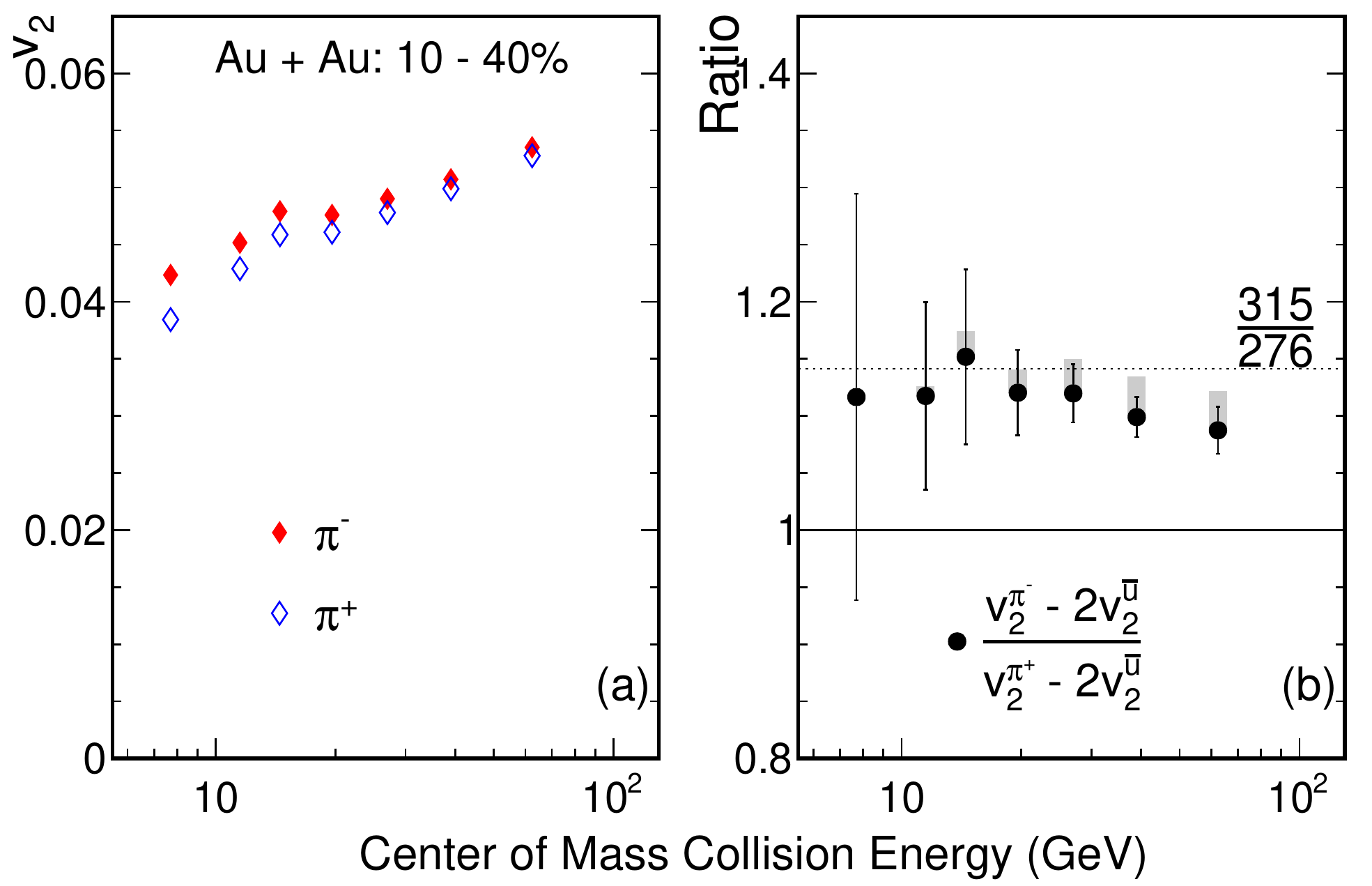}
 \caption{(Color online) (a) $v_2$ for $\pi^-$ and $\pi^+$ in 10-40\% Au+Au collisions as function of
 beam energy, based on STAR data~\cite{Olivia}. Quoted errors are statistical uncertainties only. (b) The ratio of $(v_2^{\pi^-} - 2v_2^{\bar u})$ over $(v_2^{\pi^+} - 2v_2^{\bar u})$, in comparison with the ratio between the numbers of constituent $d$ and $u$ quarks in $_{79}^{197}$Au, $315/276$. The shaded boxes represent the speculated results where the $v_2$ values for $\pi^-$ and $\pi^+$ are artificially scaled down by relative $5\%$ in view of the tracking inefficiency.
 }
 \label{fig:m5}
\end{figure}

\section{Pions and multi-strange hadrons}
Pions are the most abundant hadrons produced in heavy-ion collisions under study. $\pi^-({\bar u}d)$ and $\pi^+(u{\bar d})$ are almost symmetric, with their constituent quarks coming from similar sources: $\bar u$ and $\bar d$ quarks are produced, while $u$ and $d$ could be either produced or transported. Figure~\ref{fig:m5}(a) shows $v_2$ for $\pi^-$ and $\pi^+$ in 10-40\% Au+Au collisions as function of beam energy, based on STAR data~\cite{Olivia}.
$\pi^-$ and $\pi^+$ are close to each other at higher collision energies, and deviate towards lower energies. The ordering between $\pi^-$ and $\pi^+$ could be explained by the numbers of constituent $u$ and $d$ quarks inside the initial nuclei~\cite{Dunlop}: there are 315 $d$ quarks and only 276 $u$ quarks in $_{79}^{197}$Au. In consideration of this slight asymmetry, $v_2$ for pions can be decomposed as
\begin{eqnarray}
v_2^{\pi^-} &=& N_{{\rm trans.}d}^{\pi^-}\cdot v_2^{{\rm trans.}d} + (2-N_{{\rm trans.}d}^{\pi^-})v_2^{\bar u} \\
v_2^{\pi^+} &=& N_{{\rm trans.}u}^{\pi^+}\cdot v_2^{{\rm trans.}u} + (2-N_{{\rm trans.}u}^{\pi^+})v_2^{\bar d},
\end{eqnarray}
where $N_{{\rm trans.}d}^{\pi^-}$ and 
$N_{{\rm trans.}u}^{\pi^+}$ denote the numbers of transported  $d$ and $u$ quarks per $\pi^-$ and $\pi^+$, respectively. Still assuming $v_2^{{\rm trans.}u} = v_2^{{\rm trans.}d}$ and $v_2^{\bar u} = v_2^{\bar d}$, we reach  
\begin{equation}
\frac{N_{{\rm trans.}d}^{\pi^-}}{N_{{\rm trans.}u}^{\pi^+}} = \frac{v_2^{\pi^-}-2v_2^{\bar u}}{v_2^{\pi^+}-2v_2^{\bar u}}.
\end{equation}
Figure~\ref{fig:m5}(b) presents the ratio of $(v_2^{\pi^-} - 2v_2^{\bar u})$ over $(v_2^{\pi^+} - 2v_2^{\bar u})$, with a reasonable consistency with $315/276$. Note that pions have a lower mean $p_T$ than other hadrons, and thus are more affected by the lower $p_T$ bound of 0.2 GeV/$c$ and by the detector inefficiency, which is typically more severe at lower $p_T$. The shaded boxes represent the speculated results where the $v_2$ values for $\pi^-$ and $\pi^+$ are artificially scaled down by relative $5\%$ in view of the lower $p_T$ bound and the tracking inefficiency. In general, such a manipulation would bring up the central values of the ratios close to $315/276$. Compared with other hadrons, pions are more subject to resonance decay contributions. The decay parents tend to have higher $p_T$ and hence larger $v_2$ values, so that the daughter pions would inherit larger $v_2$ than expected by coalescence. Therefore, we refrain from deriving transported-quark $v_2$ from pion data.

\begin{figure}[t]
 \includegraphics[width=0.50\textwidth]{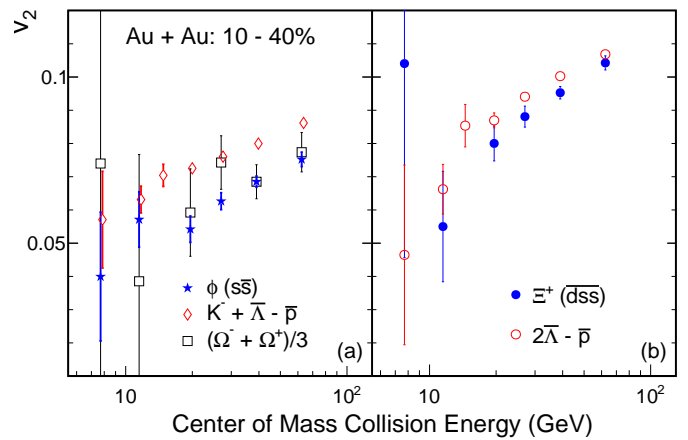}
 \caption{(Color online) $v_2$ versus beam energy in intermediate-centrality (10-40\%) Au+Au collisions, for $\phi$ mesons (a) and $\Xi^+$ hyperons (b), in comparison with the corresponding predictions of the coalescence sum rule for produced quarks, based on STAR data~\cite{Olivia}. Some data points are staggered horizontally to improve visibility.}
 \label{fig:m6}
\end{figure}

The coalescence production of multi-strange hadrons, such as $\phi$ mesons and $\Xi$ and $\Omega$ hyperons,
relies on the local strange quark density, and their formation time can be reflected in the corresponding flow measurements. Figure~\ref{fig:m6}(a) shows that $v_2$ for $\phi(s{\bar s})$ is significantly smaller than the coalescence prediction with $K^- + {\bar \Lambda} - {\bar p}$, which could be attributed to the early freeze-out of $\phi$ mesons.  At later stages of the system expansion, $s$  and $\bar s$ may have sufficiently scattered inside the QGP to acquire flow, but they have a small chance to meet again or 
meet another $\bar s$ and $s$ to coalesce into $\phi$, owing to the lower strangeness abundance at lower collision energies.
Therefore, $\phi$ mesons are not expected to flow as much as the coalescence prediction using hadrons with later formation times.
A previous STAR publication~\cite{STAR_phi_Omega} has demonstrated that the productions of $\Omega^\pm$ and $\phi$ are closely related through coalescence at BES energies.
Indeed, Fig.~\ref{fig:m6}(a) suggests that  $v_2$ values for $\phi$ can be reproduced by the prediction using $(\Omega^- + \Omega^+)/3$, especially when the statistical uncertainties are smaller, implying similar formation times for $\phi$ and $\Omega^\pm$.
The penalty due to the low strangeness abundance also applies to $\Xi^{+}({\bar  d \bar s \bar s})$, when it is compared with $2\bar \Lambda - \bar p$ in Fig.~\ref{fig:m6}(b). However, as an open-strange hadron, $\Xi^{+}$ could be formed later than $\Omega^+$ (and $\phi$), and flow closer to the coalescence prediction. 

\section{Summary}
The coalescence mechanism is regarded as important dynamics for particle yield and angular anisotropy in high-energy heavy-ion collisions. We have extensively tested the coalescence sum rule in the elliptic-flow measurements using published STAR  and ALICE data~\cite{NCQ7,Olivia}. 
The tests involving $\bar \Lambda$ and net $\Lambda$ support the coalescence sum rule at all the beam energies under study, as long as different quark species are separated.
Following the idea of differentiating produced and transported quarks~\cite{v1,Dunlop,Gang}, we have  exploited the simple coalescence framework to estimate elliptic flow for produced $u$($d$, $\bar u$, $\bar d$), $s$ and $\bar s$ quarks, as well as transported $u(d)$ quarks in 10-40\% Au+Au collisions at $\sqrt{s_{\rm NN}}=$ 7.7, 11.5, 14.5, 19.6, 27, 39 and 62.4 GeV. The speculation is confirmed that transported $u(d)$ quarks bear larger $v_2$ values than produced ones~\cite{Dunlop}. A possible breakdown of degeneracy between $s$ and $\bar s$ hints at the significant role of the associated strangeness production at lower collision energies.
Even though $\pi^-$ and $\pi^+$ are almost symmetric in production and flow, we have related their $v_2$ difference to the different numbers of constituent $u$ and $d$ quarks in a gold ion. The $v_2$ measurements of multi-strange hadrons indicate early formation times for $\phi$, $\Omega^\pm$ and $\Xi^+$. The high-statistics data from the RHIC BES-II program are anticipated to improve the precision for the analyses in this article.  Future precise  flow measurements of $D^0$ and $\Lambda_c$ will impart new insights on the coalescence dynamics of quarks with very different masses. 

\section*{Acknowledgements} \label{sec:acknowledgements}
The authors are grateful to  Declan Keane, Aihong Tang, Cheuk-Yin Wong and Manvir Grewal for fruitful discussions. In particular, we  thank Jinfeng Liao for the inspiration on Boltzmann statistics. 
This work is supported by the US Department of Energy under Grant No. DE-FG02-88ER40424.



\end{document}